\documentclass{CHEP2006}
\usepackage{url,xspace,amsmath,cite}

\author{A.~Buckley\thanks{Andy.Buckley@durham.ac.uk}, W.~J.~Stirling, M.~R.~Whalley; IPPP, Durham University, England\\
J.~M.~Butterworth, J.~Monk, E.~Nurse, B.~Waugh; University College London, England}

\title{HepData and JetWeb: HEP data archiving and model validation}

\DeclareRobustCommand{\etal}{\textit{et al.}\xspace}
\DeclareRobustCommand{\Journal}[4]{{\mbox{\emph{#1}} \textbf{{#2}}, {#3} ({#4})}}
\DeclareRobustCommand{\Rplus}{\protect\nolinebreak\hspace{-.01em}\protect\raisebox{.25ex}{\small\textbf{+}}}
\DeclareRobustCommand{\plusplus}{\Rplus\Rplus}

\DeclareRobustCommand{\CC}{C\plusplus\xspace}

\begin{document}

\maketitle


\begin{abstract}
  The CEDAR collaboration is extending and combining the JetWeb and HepData
  systems to provide a single service for tuning and validating models of
  high-energy physics processes. The centrepiece of this activity is the fitting
  by JetWeb of observables computed from Monte Carlo event generator events
  against their experimentally determined distributions, as stored in HepData.
  Caching the results of the JetWeb simulation and comparison stages provides a
  single cumulative database of event generator tunings, fitted against a wide
  range of experimental quantities. An important feature of this integration is
  a family of XML data formats, called HepML.
  
  Other aspects of the CEDAR project include building a software development
  environment for high-energy physics projects and providing an archive of HEP
  computation software. These are described elsewhere in these proceedings.
\end{abstract}

\section{Introduction}
Although the Standard Model is extraordinarily successful in describing a wide
range of phenomena, some processes cannot at present be explicitly calculated.
In particular, processes such as the study of hadronic collisions, which involve
both perturbative and non-perturbative QCD effects, are difficult to model. In
these, the final state is influenced by the parton density functions (PDFs) of
the colliding beams, by multiple interactions between partons (the ``underlying
event''), by initial and final state radiation and by the hadronisation and
decay of the outgoing partons\cite{leshouchesmcreview}. Accurate modelling of
such hadronic processes is crucial for robust interpretation of data from the
LHC. While many parts of these sub-processes can be handled by perturbative
calculations, at least in part, there is still significant need for some
phenomenological modelling, not least in matching the different sub-processes to
each other.

Generic high-energy processes are typically simulated by general purpose parton
shower Monte Carlo event generators, which dress hard process matrix elements at
a given order with the more realistic features of hadronic interactions and
fragmentation. Well-known examples of such generators are Herwig\cite{herwig6}
and Pythia\cite{pythia6}. Such generators typically introduce several free or
weakly-constrained parameters, which can only be constrained by fitting the
model predictions to the experimental data. This is far from a trivial task
since the experimental conditions vary widely, involving different beam
particles, different regions of phase space and complicated observables. The
variables may be highly correlated, so tuning a given generator to a limited set
of observables may result in non-physical predictions for observables not used
in the fit.

The CEDAR project\cite{cedar:chep04,cedar:web} exists to provide a standard,
robust and simple system for performing simultaneous data-to-model comparisons.
Its main focus is the integration of the HepData\cite{cedar:oldhepdata,
  cedar:hepdata} and JetWeb\cite{cedar:jetweb} services, improving JetWeb's
ability to constrain Monte Carlo simulation parameters. The rest of this article
will describe the projects comprising CEDAR and how this goal can be achieved.

\section{HepML}
\label{sec:hepml}
Since CEDAR's central mission is to interface HepData and JetWeb, a common
format for data record exchange is an important component. To this end, CEDAR is
defining a family of XML-based data formats, called HepML\cite{cedar:hepml},
using the XML schema\cite{schema} language. XML has been chosen because it has a
familiar plain-text representation and because it is a rapidly evolving
technology with many useful manipulation tools and libraries freely and widely
available: XML manipulation libraries are now a standard library component in
many popular modern programming languages.

At the time of writing, CEDAR HepML contains two major sub-schemas: a HepData
schema for representing HepData records (complete with meta-data and isolated
error sources) and a generator schema for representing Monte Carlo event
generator configurations. The main criticisms of XML --- that its hierarchical
data structure is restrictive and that the plain text representation is too
bulky --- are not problems for these applications, as the data involved are
naturally hierarchical with simple relational components and the quantities of
data involved are much smaller than, for example, full event records or raw
analysis data.

The family of formats as a whole, rather than any particular schema component,
is what is referred to as HepML. The intention is that CEDAR will be the curator
of the HepML family, with groups and individuals outside CEDAR being able to
propose and develop additional schemas to be incorporated into the family. The
generator sub-schema is under consideration for use as a common event generator
configuration format for the new \CC event generators.

One major benefit of using XML is that the related XPath\cite{xpath} and
XSLT\cite{xslt} technologies provide a flexible and robust system for
transforming XML documents between XML formats and into other plain text based
formats. This technology is being used to provide a variety of output modes for
HepData based on a single XML record dynamically generated from the database. In
addition to the HepML schemas, which have been released for comment and are
documented with examples at \url{http://hepforge.cedar.ac.uk/hepml/}, we intend
to release Python and Java APIs for manipulating HepML records.

We are aware of a clash between the CEDAR use of the name ``HepML'' and that
used by the CERN generator services group as part of the MCDB\cite{mcdb,mcdblh}
project. MCDB has proposed the use of an XML format for generator log files,
which is similar to certain aspects of the CEDAR HepML generator sub-schema.  We
hope that this name clash can in fact lead to a positive outcome, specifically
the development of a single suite of XML schemas for HEP applications, by
collaboration to adopt and incorporate the best features of the various
proposals when they are defined and released.


\section{HepData}
\label{sec:hepdata}
HepData is a database of general high-energy physics reaction data, and has been
maintained at Durham since the mid-1970s. It contains records from as early as
1968. The experimental data handled by HepData is that published in
peer-reviewed journal papers, and is typically collected manually by the
HepData staff, although some experiments are more pro-active in ensuring that
their data makes its way into the database. As a rule of thumb, the HepData
reaction database records scattering data such as total and differential
cross-sections, polarisation measurements and structure functions. Complementary
data such as branching ratios, $\mathcal{CP}$ asymmetries and so on are
considered the preserve of the Particle Data Group (PDG).  In addition, HepData
hosts an online parton density function (PDF) server and provides mirrors of the
SLAC Spires publications database and the Berkeley PDG website.

Here we are primarily concerned with the HepData reaction database, which is
based on the hierarchical Berkeley database management system (BDMS), and
accessed via legacy Fortran routines. This database system is now roughly 30
years old and is no longer actively maintained. It has little in the way of
modern database features such as network awareness and the central paradigm in
database systems has since shifted from strictly hierarchical databases to the
more flexible relational structure.

To make HepData suitable for remote access by JetWeb, as well as for unspecified
future uses, HepData is being migrated to a modern relational database
management system (RDBMS) with a re-designed data model. This is being
implemented via a new Java object model which reflects the structure of stored
data: published papers contain data sets, which themselves are sub-partitioned
into axes, data points and various types of error. A variety of meta-data is
stored at each level, and is used for richer querying of the database. The open
source MySQL database\cite{mysql} is being used as the RDBMS back-end, with the
coupling between the database and the Java objects to be managed via the
Hibernate persistency system\cite{hibernate}. Substantial work has been done on
migrating the database from the BDMS system, including much sanitising of the
data. The migration is an ongoing process, until the new system is declared
stable and the legacy system decommissioned: required additions to the migration
include converting the legacy data to use a unified units system.

Rather than query the database directly, users will query a Web-based front-end
which will present the data records in a choice of formats. These are foreseen
to include HTML-formatted data tables, plain text, HepML records (see
Section~\ref{sec:hepml}) and AIDA XML records\cite{aida}, with the potential for
many more. The technologies being applied here are Java servlets to provide the
database querying logic, HepML format and XSLT transformations thereof for data
transfer and presentation and Java Server Pages (JSP) for the remaining
presentation and form handling. The Java servlet and JSP execution are performed
within the Apache Tomcat servlet container, run behind the high-performance
Apache HTTPD 2 Web server. Proof of concept demonstrations of the new database
are under development on the HepData website.

An eventual aim of the HepData upgrade is to provide experiments from the LHC
era onward with a more direct way to submit their data to HepData. The HepML
format is central to this, as it is a well-structured, yet human-readable, plain
text representation of HepData records. We envisage experiments generating HepML
along with their publication plots and data tables, then submitting the HepML to
HepData using Grid authentication under the relevant experiment's virtual
organisation (VO) for checking by the HepData manager. However, such plans are
in their infancy, with the release for comment of HepML being an important first
step.

\section{JetWeb}
\label{sec:jetweb}
JetWeb is a system developed at University College London for validation of
Monte Carlo event generator tunings. Internally, JetWeb comprises a set of Java
classes which store, update and compare binned distributions. These classes are
tied to a Web interface which allows users to view the results of existing
MC-to-data comparisons and to request generation of additional simulated events
to improve the statistics associated with a given tuning. A MySQL database is
used to store observable distributions from the Monte Carlo simulations and the
Web interface is provided using the Tomcat + Apache HTTPD recipe already
described in connection with HepData.

In the first implementation of JetWeb, which is now offline, selected
experimental data is stored in a MySQL database in addition to the Monte Carlo
distributions and fit details. This is being replaced by the ability of JetWeb
to directly query HepData's records, a ``single source'' approach which benefits
JetWeb in that the potential for error when converting between HepData's
formatted data and JetWeb's database is removed and that JetWeb will
benefit automatically from any corrections to HepData's records.

A typical use of JetWeb is for a user to specify a number of generator
parameters and a number of events via the Web interface. JetWeb then determines
if Monte Carlo data is already available and distributes simulation jobs if not.
If data is available, the comparisons of MC data to experimental measurements
are displayed and the user can request more MC data to be generated if the
available statistics are judged insufficient. In the current system, JetWeb
outputs a job submission script for each data request: this must then be
submitted and the results merged by the system maintainer. Eventually, JetWeb
will use the Grid identity of the user who made the request to automatically
distribute the event generator runs.

At present, the choice of event generator parameter combinations and the
required event sample sizes are JetWeb user choices; an obvious extension of
JetWeb is to automatically sample the space of parameter combinations using e.g.
a Markov Chain Monte Carlo (MCMC) or genetic algorithm sampler and this has been
accounted for in the design of JetWeb. It may also be desirable to automate the
generation of extra events and to use the Geant Statistical Toolkit\cite{stattoolkit} for more extended statistical tests.

The predictions for observable distributions from Monte Carlo events are not
performed by the JetWeb engine itself, but by routines in the Fortran ``HZTool''
library\cite{cedar:hztool}, also maintained by CEDAR. HZTool is a library of
routines corresponding to specific experimental measurements, combined with a
selection of utility functions such as jet clustering algorithms. Each HzTool
routine roughly corresponds to a published experimental paper and as such they
tend to be provided by the primary author of the paper, in some cases outside
the CEDAR group.


As HzTool is Fortran-based and high-energy physics experimental computing has
made a definitive shift to object oriented languages, in particular \CC, CEDAR
has begun work to develop an object-oriented replacement for HZTool, titled
``Robust Validation of Experiment and Theory'' (Rivet)\cite{cedar:rivet}. As in
recent versions of HzTool, Rivet is designed to be independent of generator
details, with these being isolated into a companion package called RivetGun.
This will take a HepML generator record as an input format, translate the
appropriate model definitions into generator-specific parameters and will
transparently distribute jobs with the parameters passed to any of the supported
generators.  Both HzTool and Rivet are described in more detail elsewhere in
these proceedings\cite{cedar:rivetchep}.


\section{Conclusions}
We have described how CEDAR is combining JetWeb and HepData to provide a
definitive event generator tuning service. An important component in this effort
is the definition of the HepML family of XML data formats; these are used to
define HepData records and event generator parameters. A first version of HepML
has recently been made available for comment.

Progress has been made on both the HepData and JetWeb aspects of the CEDAR
project. The bulk of HepData has been successfully migrated from the legacy BDMS
database to the relational MySQL database, using a new Java object model. A
proof of concept demonstrator of the new HepData database, using XSLT
transformations of HepML records for data presentation, is available on the
CEDAR HepData website. JetWeb has been significantly updated to make the
addition of new event generator models much easier, to use AIDA-compliant data
plotting and to use generator schema HepML for populating its database of event
generator default parameters.

In addition, much work has been done on HzTool, HzSteer and their \CC based
replacements, and on providing HepForge\cite{cedar:hepforge,cedar:hepforgechep,cedar:hepcode}, a
lightweight development environment and repository of phenomenology programs
(described elsewhere in these proceedings). CEDAR is on track to provide robust,
globally validated event generator tunings for LHC physics analyses.


\section{Acknowledgements}
The CEDAR team would like to thank the UK Particle Physics \& Astronomy
Research Council (PPARC) for their generous support of CEDAR.

\end{document}